\catcode`\@=11					



\font\fiverm=cmr5				
\font\fivemi=cmmi5				
\font\fivesy=cmsy5				
\font\fivebf=cmbx5				

\skewchar\fivemi='177
\skewchar\fivesy='60


\font\sixrm=cmr6				
\font\sixi=cmmi6				
\font\sixsy=cmsy6				
\font\sixbf=cmbx6				

\skewchar\sixi='177
\skewchar\sixsy='60


\font\sevenrm=cmr7				
\font\seveni=cmmi7				
\font\sevensy=cmsy7				
\font\sevenit=cmti7				
\font\sevenbf=cmbx7				

\skewchar\seveni='177
\skewchar\sevensy='60


\font\eightrm=cmr8				
\font\eighti=cmmi8				
\font\eightsy=cmsy8				
\font\eightit=cmti8				
\font\eightbf=cmbx8				

\skewchar\eighti='177
\skewchar\eightsy='60


\font\ninei=cmmi9
\font\ninesy=cmsy9

\skewchar\ninei='177
\skewchar\ninesy='60


\font\tenrm=cmr10				
\font\teni=cmmi10				
\font\tensy=cmsy10				
\font\tenex=cmex10				
\font\tenit=cmti10				
\font\tensl=cmsl10				
\font\tenbf=cmbx10				
\font\tentt=cmtt10				
\font\tenss=cmss10				
\font\tensc=cmcsc10				
\font\tenbi=cmmib10				

\skewchar\teni='177
\skewchar\tenbi='177
\skewchar\tensy='60

\def\tenpoint{\ifmmode\err@badsizechange\else
	\textfont0=\tenrm \scriptfont0=\sevenrm \scriptscriptfont0=\fiverm
	\textfont1=\teni  \scriptfont1=\seveni  \scriptscriptfont1=\fivemi
	\textfont2=\tensy \scriptfont2=\sevensy \scriptscriptfont2=\fivesy
	\textfont3=\tenex \scriptfont3=\tenex   \scriptscriptfont3=\tenex
	\textfont4=\tenit \scriptfont4=\sevenit \scriptscriptfont4=\sevenit
	\textfont5=\tensl
	\textfont6=\tenbf \scriptfont6=\sevenbf \scriptscriptfont6=\fivebf
	\textfont7=\tentt
	\textfont8=\tenbi \scriptfont8=\seveni  \scriptscriptfont8=\fivemi
	\def\rm{\tenrm\fam=0 }%
	\def\it{\tenit\fam=4 }%
	\def\sl{\tensl\fam=5 }%
	\def\bf{\tenbf\fam=6 }%
	\def\tt{\tentt\fam=7 }%
	\def\ss{\tenss}%
	\def\sc{\tensc}%
	\def\bmit{\fam=8 }%
	\rm\setparameters\setbaselines\fi}


\font\twelverm=cmr12				
\font\twelvei=cmmi12				
\font\twelvesy=cmsy10	scaled\magstep1		
\font\twelveex=cmex10	scaled\magstep1		
\font\twelveit=cmti12				
\font\twelvesl=cmsl12				
\font\twelvebf=cmbx12				
\font\twelvett=cmtt12				
\font\twelvess=cmss12				
\font\twelvesc=cmcsc10	scaled\magstep1		
\font\twelvebi=cmmib10	scaled\magstep1		

\skewchar\twelvei='177
\skewchar\twelvebi='177
\skewchar\twelvesy='60

\def\twelvepoint{\ifmmode\err@badsizechange\else
	\textfont0=\twelverm \scriptfont0=\eightrm \scriptscriptfont0=\sixrm
	\textfont1=\twelvei  \scriptfont1=\eighti  \scriptscriptfont1=\sixi
	\textfont2=\twelvesy \scriptfont2=\eightsy \scriptscriptfont2=\sixsy
	\textfont3=\twelveex \scriptfont3=\tenex   \scriptscriptfont3=\tenex
	\textfont4=\twelveit \scriptfont4=\eightit \scriptscriptfont4=\sevenit
	\textfont5=\twelvesl
	\textfont6=\twelvebf \scriptfont6=\eightbf \scriptscriptfont6=\sixbf
	\textfont7=\twelvett
	\textfont8=\twelvebi \scriptfont8=\eighti  \scriptscriptfont8=\sixi
	\def\rm{\twelverm\fam=0 }%
	\def\it{\twelveit\fam=4 }%
	\def\sl{\twelvesl\fam=5 }%
	\def\bf{\twelvebf\fam=6 }%
	\def\tt{\twelvett\fam=7 }%
	\def\ss{\twelvess}%
	\def\sc{\twelvesc}%
	\def\bmit{\fam=8 }%
	\rm\setparameters\setbaselines\fi}


\font\fourteenrm=cmr12	scaled\magstep1		
\font\fourteeni=cmmi12	scaled\magstep1		
\font\fourteensy=cmsy10	scaled\magstep2		
\font\fourteenex=cmex10	scaled\magstep2		
\font\fourteenit=cmti12	scaled\magstep1		
\font\fourteensl=cmsl12	scaled\magstep1		
\font\fourteenbf=cmbx12	scaled\magstep1		
\font\fourteentt=cmtt12	scaled\magstep1		
\font\fourteenss=cmss12	scaled\magstep1		
\font\fourteensc=cmcsc10 scaled\magstep2	
\font\fourteenbi=cmmib10 scaled\magstep2	

\skewchar\fourteeni='177
\skewchar\fourteenbi='177
\skewchar\fourteensy='60

\def\fourteenpoint{\ifmmode\err@badsizechange\else
	\textfont0=\fourteenrm \scriptfont0=\tenrm \scriptscriptfont0=\sevenrm
	\textfont1=\fourteeni  \scriptfont1=\teni  \scriptscriptfont1=\seveni
	\textfont2=\fourteensy \scriptfont2=\tensy \scriptscriptfont2=\sevensy
	\textfont3=\fourteenex \scriptfont3=\tenex \scriptscriptfont3=\tenex
	\textfont4=\fourteenit \scriptfont4=\tenit \scriptscriptfont4=\sevenit
	\textfont5=\fourteensl
	\textfont6=\fourteenbf \scriptfont6=\tenbf \scriptscriptfont6=\sevenbf
	\textfont7=\fourteentt
	\textfont8=\fourteenbi \scriptfont8=\tenbi \scriptscriptfont8=\seveni
	\def\rm{\fourteenrm\fam=0 }%
	\def\it{\fourteenit\fam=4 }%
	\def\sl{\fourteensl\fam=5 }%
	\def\bf{\fourteenbf\fam=6 }%
	\def\tt{\fourteentt\fam=7}%
	\def\ss{\fourteenss}%
	\def\sc{\fourteensc}%
	\def\bmit{\fam=8 }%
	\rm\setparameters\setbaselines\fi}


\font\seventeenrm=cmr10 scaled\magstep3		


\newdimen\rp@
\newcount\@basestretchnum
\newskip\@baseskip
\newskip\headskip
\newskip\footskip


\def\setparameters{\rp@=.1em
	\headskip=24\rp@
	\footskip=\headskip
	\delimitershortfall=5\rp@
	\nulldelimiterspace=1.2\rp@
	\scriptspace=0.5\rp@
	\abovedisplayskip=10\rp@ plus3\rp@ minus5\rp@
	\belowdisplayskip=10\rp@ plus3\rp@ minus5\rp@
	\abovedisplayshortskip=5\rp@ plus2\rp@ minus4\rp@
	\belowdisplayshortskip=10\rp@ plus3\rp@ minus5\rp@
	\normallineskip=\rp@
	\lineskip=\normallineskip
	\normallineskiplimit=0pt
	\lineskiplimit=\normallineskiplimit
	\jot=3\rp@
	\setbox0=\hbox{\the\textfont3 B}\p@renwd=\wd0
	\skip\footins=12\rp@ plus3\rp@ minus3\rp@
	\skip\topins=0pt plus0pt minus0pt}


\def\setbaselines{\maxdepth=4\rp@\baselinestretch=\@basestretchnum}


\def\baselinestretch{\afterassignment\@basestretch\@basestretchnum}
\def\@basestretch{%
	\@baseskip=12\rp@ \divide\@baseskip by1000
	\normalbaselineskip=\@basestretchnum\@baseskip
	\baselineskip=\normalbaselineskip
	\bigskipamount=\the\baselineskip
		plus.25\baselineskip minus.25\baselineskip
	\medskipamount=.5\baselineskip
		plus.125\baselineskip minus.125\baselineskip
	\smallskipamount=.25\baselineskip
		plus.0625\baselineskip minus.0625\baselineskip
	\setbox\strutbox=\hbox{\vrule height.708\baselineskip
		depth.292\baselineskip width0pt }}



\def\makeheadline{\vbox to0pt{\baselinestretch=1000
	\vskip-\headskip \vskip1.5pt
	\line{\vbox to\ht\strutbox{}\the\headline}\vss}\nointerlineskip}

\def\makefootline{\baselineskip=\footskip\line{\the\footline}}

\def\big#1{{\hbox{$\left#1\vbox to8.5\rp@ {}\right.\n@space$}}}
\def\Big#1{{\hbox{$\left#1\vbox to11.5\rp@ {}\right.\n@space$}}}
\def\bigg#1{{\hbox{$\left#1\vbox to14.5\rp@ {}\right.\n@space$}}}
\def\Bigg#1{{\hbox{$\left#1\vbox to17.5\rp@ {}\right.\n@space$}}}


\mathchardef\alpha="710B
\mathchardef\beta="710C
\mathchardef\gamma="710D
\mathchardef\delta="710E
\mathchardef\epsilon="710F
\mathchardef\zeta="7110
\mathchardef\eta="7111
\mathchardef\theta="7112
\mathchardef\iota="7113
\mathchardef\kappa="7114
\mathchardef\lambda="7115
\mathchardef\mu="7116
\mathchardef\nu="7117
\mathchardef\xi="7118
\mathchardef\pi="7119
\mathchardef\rho="711A
\mathchardef\sigma="711B
\mathchardef\tau="711C
\mathchardef\upsilon="711D
\mathchardef\phi="711E
\mathchardef\chi="711F
\mathchardef\psi="7120
\mathchardef\omega="7121
\mathchardef\varepsilon="7122
\mathchardef\vartheta="7123
\mathchardef\varpi="7124
\mathchardef\varrho="7125
\mathchardef\varsigma="7126
\mathchardef\varphi="7127
\mathchardef\imath="717B
\mathchardef\jmath="717C
\mathchardef\ell="7160
\mathchardef\wp="717D
\mathchardef\partial="7140
\mathchardef\flat="715B
\mathchardef\natural="715C
\mathchardef\sharp="715D


\def\err@badsizechange{%
	\immediate\write16{--> Size change not allowed in math mode, ignored}}

\baselinestretch=1000
\tenpoint

\catcode`\@=12					

\catcode`\@=11
\expandafter\ifx\csname @iasmacros\endcsname\relax
	\global\let\@iasmacros=\par
\else	\immediate\write16{}
	\immediate\write16{Warning:}
	\immediate\write16{You have tried to input iasmacros more than once.}
	\immediate\write16{}
	\endinput
\fi
\catcode`\@=12


\def\rmb{\seventeenrm}

\def\singlespace{\baselineskip=\normalbaselineskip}
\def\halfspace{\baselineskip=1.5\normalbaselineskip}
\def\doublespace{\baselineskip=2\normalbaselineskip}


\def\AB{\bigskip\parindent=40pt
        \centerline{\bf ABSTRACT}\medskip\halfspace\narrower}
\def\AE{\bigskip\nonarrower\doublespace}
\def\nonarrower{\advance\leftskip by-\parindent
	\advance\rightskip by-\parindent}


\def\boxit#1{\vbox{\hrule\hbox{\vrule\kern3pt
	\vbox{\kern3pt#1\kern3pt}\kern3pt\vrule}\hrule}}

\def\hence{\leavevmode\hbox{\bf .\raise5.5pt\hbox{.}.} }

\def\dalemb#1#2{{\vbox{\hrule height.#2pt
	\hbox{\vrule width.#2pt height#1pt \kern#1pt \vrule width.#2pt}
	\hrule height.#2pt}}}
\def\gtorder{\mathrel{\raise.3ex\hbox{$>$}\mkern-14mu
             \lower0.6ex\hbox{$\sim$}}}
\def\ltorder{\mathrel{\raise.3ex\hbox{$<$}\mkern-14mu
             \lower0.6ex\hbox{$\sim$}}}

\newdimen\fullhsize
\newbox\leftcolumn
\def\twoup{\hoffset=-.5in \voffset=-.25in
  \hsize=4.75in \fullhsize=10in \vsize=6.9in
  \def\fullline{\hbox to\fullhsize}
  \let\lr=L
  \output={\if L\lr
        \global\setbox\leftcolumn=\columnbox\global\let\lr=R \advancepageno
      \else \doubleformat \global\let\lr=L\fi
    \ifnum\outputpenalty>-20000 \else\dosupereject\fi}
  \def\doubleformat{\shipout\vbox{
    \fullline{\box\leftcolumn\hfil\columnbox}\advancepageno}}
  \def\columnbox{\leftline{\vbox{\makeheadline\pagebody\makefootline}}}
  \tolerance=1000 }

\twelvepoint
\doublespace
{\nopagenumbers{
\rightline{IASSNS-HEP-97/16}
\rightline{~~~March, 1997}
\bigskip\bigskip
\centerline{\rmb Poincar\'e Supersymmetry Representations Over} 
\centerline{\rmb Trace Class Noncommutative Graded Operator Algebras}
\medskip
\centerline{\it Stephen L. Adler
}
\centerline{\bf Institute for Advanced Study}
\centerline{\bf Princeton, NJ 08540}
\medskip
\bigskip\bigskip
\leftline{\it Send correspondence to:}
\medskip
{\singlespace\leftline{Stephen L. Adler}
\leftline{Institute for Advanced Study}
\leftline{Olden Lane, Princeton, NJ 08540}
\leftline{Phone 609-734-8051; FAX 609-924-8399; email adler@sns.ias.
edu}}
\bigskip\bigskip
}}
\vfill\eject
\pageno=2
\AB
We show that rigid supersymmetry theories in four dimensions can be 
extended to give supersymmetric trace (or generalized quantum) dynamics
theories, in which the supersymmetry algebra is represented by the 
generalized Poisson bracket of trace supercharges, constructed from 
fields that form a trace class noncommutative graded operator algebra.  
In particular, supersymmetry theories can be turned into 
supersymmetric matrix models this way.  We demonstrate our results by 
detailed component field calculations for the Wess-Zumino and the 
supersymmetric Yang-Mills models (the latter with axial gauge fixing), 
and then show that they are also implied by a simple and general superspace 
argument.  

\AE
\bigskip\bigskip
\vfill\eject
\pageno=3
\centerline{{\bf 1.~~Introduction to Trace Dynamics}}
In constructing supersymmetric field theories, one usually verifies the 
supersymmetry by doing a {\it classical} Grassmann calculation, treating the 
bosonic fields as classical (rather than operator) variables and the 
fermionic fields as classical Grassmann (rather than operator Grassmann) 
variables.  Then one quantizes by replacing the classical Poisson or Dirac  
brackets by commutators/anticommutators.  We shall show in this paper 
that for rigid supersymmetry theories,
a significant generalization of this standard approach is possible, in 
which the pre-quantum bosonic and fermionic fields are respectively trace 
class 
even and odd grade {\it operators}, such as, for example, $N \times N$ 
matrices 
whose matrix elements are respectively the even and odd grade elements of a 
complex Grassmann algebra.  In particular, our results show that rigid 
supersymmetry theories can be extended to give supersymmetric matrix models.  
The requirement that the field variables be of trace class is crucial to our 
results, since in the calculations given below, cyclic permutation of 
operator variables under the trace provides the necessary commutativity,  
generalizing the trivial commutativity/anticommutativity of classical 
field variables, for verifying both supersymmetry of the Lagrangian and 
the closure of the supersymmetry algebra.  

Our approach is based on the trace (or generalized quantum) dynamics that 
we have proposed [1] and studied with various collaborators [2]; we in fact 
shall use a simplified form of this dynamics that becomes possible when 
Grassmann algebras are employed to represent the fermion/boson distinction.  
Let $B_1$ and $B_2$ be two $N \times N$ matrices with matrix elements that 
are even grade elements of a complex Grassmann algebra, and Tr the ordinary 
matrix trace, which obeys the cyclic property 
$${\rm Tr} B_1 B_2 = \sum_{m,n}(B_1)_{mn}(B_2)_{nm} 
=\sum_{m,n} (B_2)_{nm}(B_1)_{mn}= {\rm Tr}B_2B_1~~~.
\eqno(1a)$$
Similarly, let $\chi_1$ and $\chi_2$ be two $N \times N$ matrices with 
matrix elements that are odd grade elements of a complex Grassmann algebra, 
which anticommute rather than commute, so that the cyclic property for these 
takes the form 
$${\rm Tr}\chi_1 \chi_2 = \sum_{m,n} (\chi_1)_{mn}(\chi_2)_{nm}
=-\sum_{m,n}(\chi_2)_{nm}(\chi_1)_{mn}=-{\rm Tr}\chi_2\chi_1~~~.
\eqno(1b)$$
The cyclic/anticyclic properties of Eqs.~(1a, 1b) are just those assumed  
for the trace operation {\bf Tr} of trace dynamics, although in Refs. [1, 2] 
the fermionic operators were realized as matrices with complex matrix 
elements, all of which anticommute with a grading operator $(-1)^F$ which 
formed part of the definition of {\bf Tr}.  Since the use of Grassmann odd 
fermions eliminates the need for the inclusion of the $(-1)^F$ factor
\footnote{*} {If the $(-1)^F$ construction is combined with Grassmann odd 
fermions one gets  the ``supertrace'' str, that obeys the cyclic property  
$str N_1 N_2 = str N_2 N_1$ for both bosonic and fermionic $N_{1,2}$.  We 
will not use the supertrace in this article. }, we shall continue here to  
use the notation Tr, with the understanding that fermionic matrices or 
operators obey Eq.~(1b) while bosonic matrices or operators obey Eq.~(1a).  
From Eqs.~(1a, b), one immediately derives the trilinear cyclic identities 
$$\eqalign{
{\rm Tr} B_1[B_2,B_3]=&{\rm Tr}B_2[B_3,B_1]={\rm Tr}B_3[B_1,B_2]~~~\cr
{\rm Tr} B_1\{B_2,B_3\}=&{\rm Tr}B_2\{B_3,B_1\}={\rm Tr}B_3\{B_1,B_2\}  \cr
{\rm Tr} B \{\chi_1,\chi_2\}=&{\rm Tr}\chi_1 [\chi_2,B]
={\rm Tr} \chi_2[\chi_1,B]  \cr
{\rm Tr} \chi_1\{B,\chi_2\}=&{\rm Tr}\{\chi_1,B\}\chi_2
={\rm Tr}[\chi_1,\chi_2]B ~~~, \cr
}\eqno(1c) $$
which are repeatedly used below.  

The basic observation of trace dynamics is that given the trace of a 
polynomial $P$ constructed from noncommuting matrix or operator variables, 
one can define a derivative of the $c$-number ${\rm Tr} P$ with respect to 
an operator variable ${\cal O}$ by varying and then cyclically permuting 
so that in each term the factor $\delta {\cal O}$ stands on the right, 
giving the fundamental definition 
$$\delta {\rm Tr P}={\rm Tr} {\delta {\rm Tr}P \over \delta {\cal O}} 
\delta {\cal O}~~~,\eqno(2a)$$
or in the condensed notation that we shall use throughout this paper, in 
which ${\bf P} \equiv {\rm Tr}P$, 
$$\delta {\bf P}   = {\rm Tr} {\delta {\bf P} \over \delta {\cal O}}
\delta {\cal O}~~~.\eqno(2b)$$
Letting ${\bf L}[\{q_r\},\{\dot q_r\}]$ be a trace Lagrangian that is a 
function of the bosonic or fermionic operators $\{q_r\}$ and their time 
derivatives, and requiring that the trace 
action ${\bf S}=\int dt {\bf L}$ be stationary with respect to variations 
of the $q_r$'s that preserve their bosonic or fermionic type, 
one finds [1] the operator 
Euler-Lagrange equations 
$${\delta {\bf L} \over \delta q_r} -
{d \over dt} {\delta {\bf L} \over \delta \dot q_r} =0~~~.\eqno(2c)$$
Defining the momentum operator $p_r$ conjugate to $q_r$, which is of the 
same bosonic or fermionic type as $q_r$, by 
$$p_r \equiv {\delta {\bf L} \over \delta \dot q_r}~~~,\eqno(3a)$$
the trace Hamiltonian {\bf H} is defined by 
$${\bf H}={\rm Tr}\sum_rp_r \dot q_r - {\bf L}~~~.\eqno(3b)$$
Performing general same-type operator variations, and using the Euler-Lagrange 

equations, we find from Eq.~(3b) that the trace Hamiltonian {\bf H} is a 
trace functional of the operators $\{q_r\}$ and $\{p_r\}$, 
$${\bf H}= {\bf H}[\{q_r\},\{p_r\}]~~~,\eqno(4a)$$
with the operator derivatives 
$${\delta {\bf H} \over \delta q_r}=-\dot p_r~,~~~
{\delta {\bf H} \over \delta p_r}=\epsilon_r \dot q_r~,~~~\eqno(4b)$$
with $\epsilon_r=1(-1)$ according to whether $q_r,p_r$ are bosonic 
(fermionic).  
Letting {\bf A} and {\bf B} be two trace functions of the operators 
$\{q_r\}$ and $\{p_r\}$, it is convenient to define the {\it generalized 
Poisson bracket} 
$$\{{\bf A}, {\bf B} \}={\rm Tr} \sum_r \epsilon_r \left(
{\delta {\bf A} \over \delta q_r}{\delta {\bf B} \over \delta p_r}
-{\delta {\bf B} \over \delta q_r} {\delta {\bf A} \over \delta p_r} \right)
~~~.\eqno(5a)$$
Then using the Hamiltonian form of the equations of motion, one readily 
finds that for a general trace functional ${\bf A}[\{q_r\},\{p_r\}]$, 
the time derivative is given by 
$${d \over dt} {\bf A}=\{ {\bf A}, {\bf H} \}~~~;\eqno(5b)$$
in particular, letting {\bf A} be the trace Hamiltonian {\bf H}, and using 
the fact that the generalized Poisson bracket is antisymmetric in its 
arguments, it follows that the time derivative of {\bf H} vanishes.  

An important property of the generalized Poisson bracket is that it 
satisfies [2] the Jacobi identity, 
$$\{ {\bf A},\{ {\bf B},{\bf C} \}\}    
+\{ {\bf C},\{ {\bf A},{\bf B} \}\}    
+\{ {\bf B},\{ {\bf C},{\bf A} \}\} =0~~~.\eqno(6a)$$   
As a consequence, if ${\bf Q_1}$ and ${\bf Q_2}$ are two conserved charges, 
that is if
$$0={d \over dt} {\bf Q_1} = \{ {\bf Q_1}, {\bf H} \}~,~~~ 
  0={d \over dt} {\bf Q_2} = \{ {\bf Q_2}, {\bf H} \}~,~~~ 
  \eqno(6b)$$ 
then their generalized Poisson bracket $\{ {\bf Q_1}, {\bf Q_2} \}$ 
also has a vanishing generalized Poisson bracket with {\bf H}, and is 
conserved.  This is how we will use the trace dynamics formalism to 
get representations of the Poincar\'e 
supersymmetry algebra in the following sections.  

A significant feature of trace dynamics is that, as discovered by 
Millard [3], the operator [3, 4] 
$$\tilde C \equiv \sum_{r~{\rm bosons}}[q_r,p_r]-\sum_{r~{\rm fermions}}
\{q_r, p_r\}~~~\eqno(7)$$
is conserved by the dynamics.  Making the assumption (which may presuppose  
taking the $N \rightarrow \infty $ limit) that trace dynamics is ergodic, 
one can then analyze [4] the statistical mechanics of trace dynamics for the 
generic case in which the conserved quantities are the trace Hamiltonian 
{\bf H} and the operator $\tilde C$.  In the analysis as given in [4], the 
realization of fermions using the $(-1)^F$ construction explicitly entered 
the argument in two places.  The first was in the demonstration that  
trace dynamics has a generalized Liouville theorem, which is the foundation 
for a statistical treatment.  It is easy to see that this demonstration 
remains valid when the fermions are realized by Grassmann odd matrices 
without use of the $(-1)^F$ construction.\footnote{*}{  In the  
argument at the top of p. 227 of [4], a fermionic $\epsilon_r=-1$ 
was absorbed through the relation $\epsilon_r = \epsilon_m \epsilon_n$, 
with $\epsilon_{m,n}$ the state grading factors introduced by $(-1)^F$. 
When the fermions are realized through Grassmann matrices  
without the $(-1)^F$, the state grading factors are unity and are absent, but 
the interchange of the order of derivatives with respect to Grassmann odd 
matrix elements introduces an extra minus sign in the fermionic case,  again 
absorbing the fermionic $\epsilon_r =-1$ and showing that the deviation 
of the Jacobian from unity vanishes.}  The second place where the $(-1)^F$  
construction played a role was in the issue of convergence of the partition 
function; because ${\rm Tr} (-1)^F H$ is indefinite even when the operator 
Hamiltonian $H$ has a positive definite bosonic part, it was necessary in 
[4] to restrict the analysis to theories in which $Tr H$ and $Tr (-1)^F H$ 
both generated the same Hamilton equations of motion, and this led to a 
doubling 
of the complexity of the statistical analysis.  With Grassmann fermions 
this problem is avoided, since for the typical models we are studying the 
bosonic part of $H$ is a positive operator, from which  {\bf Tr H}  inherits 
good positivity properties, and so the partition function can be expected 
to converge. The canonical ensemble then takes the simple form given 
in Eq.~(48c) of [4], 
$$\eqalign{
\rho=&Z^{-1} \exp(-{\rm Tr} \tilde \lambda \tilde C - \tau {\bf H})    \cr
Z=&\int d\mu \exp(-{\rm Tr} \tilde \lambda \tilde C - \tau {\bf H})~~~,\cr
}\eqno(8)$$
with $d \mu$  the invariant matrix (or operator) phase space measure 
provided by Liouville's theorem, rather than the more complicated form 
given in Eq.~(F.1) of [4]. (As shown in [5], this canonical ensemble can  
also be derived from the corresponding microcanonical ensemble.)   
The structure of the Ward or equipartition theorems of [4] is correspondingly 
simplified, and leads as before to the conclusion that {\it the statistical 
mechanics of trace dynamics is complex quantum field theory}, with the 
average of the operator $\tilde C$ playing the role of $i \hbar$.  As 
suggested in [4], this means that trace dynamics behaves as a pre-quantum 
mechanics, in which it is likely that the ultraviolet divergences of 
quantum field theory are absent.  Corrections to the  
quantum field theory approximation are expected to be of order $\omega 
\tau$, with $\omega$ a characteristic frequency of the physics in question, 
and so we expect the inverse of the parameter  $\tau$ appearing in the 
canonical 
ensemble of Eq.~(8), which has the dimension of mass, to play a role 
analogous to that of the string tension in string theories.  
\bigskip    
\centerline{{\bf 2.~~The Wess-Zumino Model}}

We begin our discussion of component field supersymmetric models with the 
Wess-Zumino model.  We follow the notational conventions of West [6], 
except that we normalize the fermion terms in the action differently, and 
we always use the Majorana representation for the Dirac gamma matrices.  
Our explicit choice of $\gamma$ matrices is given in the Appendix, where 
we discuss the properties of {\it representation covariant} $\gamma$ matrix 
identities that take a particularly simple form when expressed in Majorana 
representation; these will play a significant role in our analysis.  

We start from the trace Lagrangian 
$$\eqalign{
{\bf L}=&\int d^3x {\rm Tr}\big( -{1\over 2} (\partial_{\mu}A)^2 -{1 \over 2} 
(\partial_{\mu} B)^2-\bar{\chi} \gamma^{\mu} \partial_{\mu} \chi + 
{1 \over 2} F^2 + {1 \over 2} G^2   \cr
&-m (AF+BG - \bar{\chi} \chi)   \cr             
&-\lambda[(A^2-B^2)F +G\{A,B\} 
-2 \bar{\chi}(A-i\gamma_5 B) \chi] \big)~~~,\cr
}\eqno(9a)$$
with $A,B,F,G$ self-adjoint bosonic $N \times N$ matrices (or operators) and 
with  
$\chi$ a fermionic 4 component column vector spinor, each spin component 
of which is a self-adjoint fermionic $N \times N$ matrix (or operator).  
The notation $\bar{\chi}$ is defined by $\bar{\chi}=\chi^T \hat\gamma^0$, 
with the transpose $T$ acting only on the Dirac spinor structure, 
so that $\chi^T$ is the 4 component row vector spinor constructed from the 
same $N \times N$ matrices that appear in $\chi$,  and $\hat\gamma^0$ is 
an abbreviation for $i \gamma^0$.  The numerical parameters $\lambda$ and 
$m$ are respectively the coupling constant and mass.  Equation (9a) is 
identical in appearance to the usual Wess-Zumino model Lagrangian, except 
that we have explicitly symmetrized the term $G\{A,B\}$; symmetrization of 
the other terms is automatic (up to total derivatives that do not contribute 
to the action) by virtue of the cyclic property of the trace.  

Taking operator variations of Eq.~(9a) by using the recipe of Eqs.~(2a, b), 
the 
Euler-Lagrange equations of Eq.~(2c) take the form
$$\eqalign{
\partial^2A=&mF+\lambda (\{A,F\}+\{B,G\}-2\bar{\chi} \chi)  \cr
\partial^2B=&mG+\lambda(-\{B,F\}+\{A,G\}+2i\bar{\chi}\gamma_5\chi)   \cr
\gamma^{\mu}\partial_{\mu}\chi=&m\chi+\lambda(\{A,\chi\}-i\{B,\gamma_5\chi\})  
\cr
F=&mA+\lambda(A^2-B^2)  \cr
G=&mB+ \lambda\{A,B\}~~~.  \cr
}\eqno(9b)$$
Transforming to Hamiltonian form, the canonical momenta of Eq.~(3a) are 
$$\eqalign{
p_{\chi}=&-\bar{\chi} \gamma^0=i\chi^T  \cr
p_A=&\partial_0 A \cr
p_B=&\partial_0 B~~~, \cr 
}\eqno(10a)$$
and the trace Hamiltonian is given by 
$$\eqalign{
{\bf H}=&\int d^3x {\rm Tr} \big({1\over 2}[p_A^2+p_B^2+(\vec \nabla A)^2
+(\vec \nabla B)^2] -ip_{\chi} \hat \gamma^0 \vec \gamma \cdot \vec \nabla 
\chi   \cr
+&{1\over 2}(F^2+G^2) 
-m\bar{\chi} \chi+i \lambda p_{\chi} \hat\gamma^0\{(A-i\gamma_5B),\chi\} 
\big)~~~, \cr
}\eqno(10b)$$
in which $F$ and $G$ are understood to be the functions of $A$ and $B$ 
given by the final two lines of Eq.~(9b), and where we have taken care   
to write {\bf H} so that it is manifestly symmetric 
in the identical quantities $p_{\chi}$ and $i\chi^T$.
The trace three-momentum $\vec{\bf P}$ is given by 
$$\vec{\bf P}=-\int d^3x {\rm Tr} (p_A \vec\nabla A+p_B \vec \nabla B
+p_{\chi} \vec \nabla \chi)~~~,\eqno(10c)$$
while the conserved operator $\tilde C$ of Eq.~(7) is given by 
$$\tilde C=\int d^3x ([A,p_A]+[B,p_B]-\{\chi,p_{\chi} \})  ~~~,\eqno(10d)$$
with a contraction of the spinor indices in the final term of Eq.~(10d) 
understood.  

Let us now perform a supersymmetry variation of the fields given by 
$$\eqalign{
\delta A=&\bar \epsilon \chi ~~~~~~~\delta B=i \bar \epsilon \gamma_5 \chi  
\cr
\delta \chi=&{1\over 2}[F+i\gamma_5 G+\gamma^{\mu}\partial_{\mu}(A+i\gamma_5 
B)]
\epsilon  \cr
\delta F=&\bar \epsilon \gamma^{\mu} \partial_{\mu} \chi~~~~~~~
\delta G=i \bar \epsilon \gamma_5 \gamma^{\mu} \partial_{\mu} \chi~~~,\cr
}\eqno(11)$$
with $\epsilon$ a $c$-number Grassmann spinor (i.e., a four component spinor, 
the spin components of which are $1 \times 1$ Grassmann matrices).
Substituting Eq.~(11) into the trace Lagrangian of Eq.~(9a), a lengthy 
calculation shows that when $\epsilon$ is constant, the variation of {\bf L} 
vanishes.   The calculation parallels that done in the conventional 
$c$-number Lagrangian case, except that the trilinear 
cyclic identities of Eq.~(1c) 
are used extensively in place of commutativity/anticommutativity of the 
fields, and the vanishing of the terms cubic in $\chi$ is most 
easily established  by using the cyclic property of the trace, which 
implies that 
$${\rm Tr}\epsilon_c \chi_a \chi_b  \chi_d ={\rm Tr} \epsilon_c \chi_d 
\chi_a \chi_b = {\rm Tr} \epsilon_c \chi_b \chi_d \chi_a~~~,\eqno(12a)$$ 
together with the cyclic identity valid for Majorana representation 
$\gamma$ matrices (see the Appendix), 
$$\sum_{{\rm cycle}~ a \rightarrow b \rightarrow d \rightarrow a} 
[\hat\gamma^0_{ab} \hat\gamma^0_{cd} + (\hat \gamma^0 \gamma_5)_{ab}
(\hat \gamma^0 \gamma_5)_{cd}]=0~~~.\eqno(12b)$$
When $\epsilon$ is not constant, the variation of {\bf L} is given by 
$$\eqalign{
\delta {\bf L}=&\int d^3x {\rm Tr}(\bar J^{\mu} \partial_{\mu} \epsilon)  \cr
\bar J^{\mu}=&-\bar \chi \gamma^{\mu}
\big[(\gamma^{\nu}\partial_{\nu}+m)
(A+i\gamma_5 B)+\lambda(A^2-B^2+i\gamma_5\{A,B\}) \big]~~~,   \cr
}\eqno(13a)$$
which identifies the trace supercharge ${\bf Q}_{\alpha}$ as 
$$\eqalign{
{\bf Q}_{\alpha}\equiv&\int d^3x {\rm Tr} \bar J^0 \alpha \cr
=&\int d^3x {\rm Tr}{1 \over 2}(p_{\chi}+i\chi^T)
\big[(\gamma^{\nu}\partial_{\nu}+m)
(A+i\gamma_5 B)+\lambda(A^2-B^2+i\gamma_5\{A,B\}) \big]\alpha ~~~,   \cr
}\eqno(13b)$$
where we have again taken care to express ${\bf Q}_{\alpha}$ 
symmetrically in the identical quantities $p_{\chi}$ and $i\chi^T$. It  
is straightforward to check, using the equations of motion and the 
cyclic identity, that 
${\rm Tr} \bar J^{\mu} $is a conserved trace supercurrent, which implies 
that the trace supercharge is conserved.  

We are now ready to check the closure of the supersymmetry algebra under 
the generalized Poisson bracket of Eq.~(5a), which for the 
Hamiltonian dynamics of the Wess-Zumino model gives
$$
\{ {\bf Q}_{\alpha},{\bf Q}_{\beta} \}={\rm Tr}\big[
{\delta {\bf Q}_{\alpha} \over \delta A} {\delta {\bf Q}_{\beta} 
\over \delta p_A}+
{\delta {\bf Q}_{\alpha} \over \delta B} {\delta {\bf Q}_{\beta} 
\over \delta p_B}
-{\delta {\bf Q}_{\alpha} \over \delta \chi} 
{\delta {\bf Q}_{\beta} 
\over \delta p_{\chi}}
-\big(\alpha \leftrightarrow \beta\big) 
\big] ~~~.\eqno(14a)$$
There are two strategies for carrying out the considerable amount of  
algebra involved in evaluating Eq.~(14a).  The first is to directly rearrange
into the expected form, verifying along the way various Majorana 
representation 
$\gamma$ matrix identities that are needed; the second is to first Fierz 
transform so as to isolate a factor of the form $\alpha^T \Gamma \beta$, 
and then to show that this yields the expected result.  We shall 
use the first method here, and the second method in discussing the 
supersymmetric Yang-Mills model in the next section.  Proceeding by the 
first method, we find that Eq.~(14) rearranges, using the cyclic identities 
of Eq.~(1c), into the form 
$$\{{\bf Q}_{\alpha},{\bf Q}_{\beta} \}=\bar\alpha \gamma^0 \beta {\bf H}
-\bar \alpha \vec \gamma \beta \cdot \vec {\bf P} ~~~,\eqno(14b)$$
with {\bf H} and $\vec {\bf P}$ the trace Hamiltonian and three-momentum
given above.  The $\gamma$ matrix identities needed can be obtained by 
repeated applications either of the cyclic identity of Eq.~(12b), 
or of the additional 
identity (with $\ell,m,n$ spatial indices, and $\epsilon_{\ell m n}$ 
the three index antisymmetric tensor with $\epsilon_{123}=1)$
$$\eqalign{
&\gamma^{\ell}_{ab}\hat \gamma^0_{cd}+\gamma^{\ell}_{db}\hat \gamma^0_{ca}
-(\gamma^{\ell}\gamma_5)_{ab}(\hat \gamma^0 \gamma_5)_{cd} 
-(\gamma^{\ell} \gamma_5)_{db}(\hat \gamma^0 \gamma_5)_{ca}  \cr
&=\delta_{ad}(\hat \gamma^0 \gamma_{\ell})_{bc}
-(\hat \gamma^0 \gamma_{\ell})_{ad} \delta_{bc} 
+\epsilon_{\ell mn}(\gamma_{\ell}\gamma_m\gamma_5)_{ad}
(\gamma_{\ell}\gamma_n)_{cb}  ~~~,\cr
}\eqno(15)$$
which we have verified by the method described in the Appendix.  
It is also easy to check that ${\bf Q}_{\epsilon}$ plays the role of the 
generator of supersymmetry transformations for the dynamical variables 
$A,B,\chi$ under the generalized Poisson bracket, since we readily find 
(for constant Grassmann even parameters $a,b$ and Grassmann odd 
parameter $c$)
$$\{ {\rm Tr}(aA+bB+c\chi), {\bf Q}_{\epsilon}\}=
{\rm Tr}(a \delta A+ b \delta B+c \delta \chi)~~~,\eqno(16)$$
with $\delta A, \delta B, \delta \chi$ the supersymmetry variations 
given by Eq.~(11) above, after 
elimination of the auxiliary fields $F,G$ by their equations of motion.

\bigskip
\centerline{\bf 3.~~The Supersymmetric Yang-Mills Model }

As our next example of a component field supersymmetric model, we discuss 
supersymmetric Yang-Mills theory.  (In Ref. [7] we have given 
a simpler analog of this discussion, in the context of the matrix model 
for M theory.)  We start from the trace Lagrangian 
$${\bf L}=\int d^3x {\rm Tr}\big[ {1\over 4g^2} F^2_{\mu\nu} -\bar \chi
\gamma^{\mu}D_{\mu} \chi +{1\over 2} D^2\big]~~~,\eqno(17a)$$
with the field strength $F_{\mu\nu}$ and covariant derivative $D_{\mu}$ 
constructed from the gauge potential $A_{\mu}$ according to 
$$\eqalign{
F_{\mu\nu}=&\partial_{\mu}A_{\nu}-\partial_{\nu}A_{\mu}+[A_{\mu},A_{\nu}]
 \cr
D_{\mu}{\cal O}=&\partial_{\mu}{\cal O}+[A_{\mu},{\cal O}] \cr
\Rightarrow&D_{\mu}F_{\nu\lambda}+D_{\nu}F_{\lambda\mu}+
D_{\lambda}F_{\mu\nu}             
=0  ~~~.\cr
}\eqno(17b)$$
In Eq.~(17b), the potential components $A_{\mu}$ are each an 
anti-self-adjoint, and 
the auxiliary field $D$ a self-adjoint, 
bosonic $N \times N$ matrix (or operator), and each spinor component of  
$\chi$ is a self-adjoint fermionic $N \times N$ matrix (or operator).
The Euler-Lagrange equations of motion are 
$$\eqalign{
D=&0   \cr
\gamma^{\mu}D_{\mu} \chi=&0   \cr
D_{\mu}F^{\mu\nu}=&2g^2\bar \chi \gamma^{\nu} \chi ~~~;\cr
}\eqno(18a)$$
as usual for a gauge system, the $\nu=0$ component of Eq.~(18a) is not  
a dynamical evolution equation, but rather the constraint
$$D_{\ell}F^{\ell 0}=2g^2 \bar \chi \gamma^0 \chi~~~.\eqno(18b)$$

Going over to the Hamiltonian formalism, the canonical momenta are 
given by 
$$p_{A_{\ell}}=-{1 \over g^2} F_{0\ell}~,~~~p_{\chi}=i\chi^T~~~, 
\eqno(19a)$$
and the axial gauge trace Hamiltonian 
(see [1] for a derivation and references) is 
$${\bf H}={\bf H}_{A} +{\bf H}_{\chi}~~~,\eqno(19b)$$
with 
$$\eqalign{
{\bf H}_{A}=&\int d^3x {\rm Tr} \big( {-g^2 \over 2} \sum_{\ell=1}^2 
p^2_{A_{\ell}} -{1 \over 2 g^2} F_{03}^2 \cr
&-{1 \over 2g^2}(\partial_1A_2-\partial_2A_1+ [A_1,A_2])^2 -{1 \over 2g^2}
[(\partial_3 A_1)^2 + (\partial_3 A_2)^2 ]\big) \cr
F_{03}=&{1\over2} g^2 \int_{-\infty}^{\infty}dz^{\prime} 
\epsilon(z-z^{\prime})
[-(p_{\chi}\chi+\chi^Tp^T_{\chi})+D_1p_{A_1}+D_2p_{A_2})\vert_{z^{\prime}}\cr
{\bf H}_{\chi}=&-i\int d^3x {\rm Tr}(p_{\chi} \hat \gamma^0\gamma_{\ell} 
D_{\ell}\chi) ~~~,\cr
}\eqno(19c)$$
where we have taken care to write {\bf H} in a form symmetric in the 
identical quantities $p_{\chi}$ and $i\chi^T$, and where $\epsilon(z)=1(-1)$  
for $z>0(z<0)$.
The trace three momentum is 
$$ {\bf P}_m=-\int d^3x {\rm Tr} (\sum_{\ell=1}^3 F_{m\ell}p_{A_{\ell}}
+p_{\chi}D_m\chi )~~~,\eqno(20a)$$
and the conserved operator $\tilde C$ of Eq.~(7) is given by 
$$\tilde C=\int d^3x (\sum_{\ell=1}^2[A_{\ell},p_{A_{\ell}}]
-\{\chi,p_{\chi}\})
~~~,\eqno(20b)$$
with a contraction of the spinor indices in the final term of Eq.~(20b) 
understood.  By virtue of the constraint of Eq.~(18b), the conserved 
operator $\tilde C$ can also be written as 
$$\tilde C=-\int d^3x \sum_{\ell=1}^2  \partial_{\ell} p_{A_{\ell}}=
-\int_{\rm sphere~at~\infty} d^2 S_{\ell}~p_{A_{\ell}}~~~,\eqno(20c)$$
which vanishes when the surface integral in Eq.~(20c) is zero.  

Making now the supersymmetry variations 
$$\eqalign{
\delta A_{\mu}=&ig \bar \epsilon \gamma_{\mu} \chi \cr
\delta \chi=&\big({i\over 8g} [\gamma_{\mu},\gamma_{\nu}]F^{\mu\nu} 
+{i \over 2} \gamma_5 D)\epsilon \cr 
\delta D=&i \bar \epsilon \gamma_5 \gamma^{\mu} D_{\mu} \chi~~~,  \cr
}\eqno(21a)$$
in the trace Lagrangian, we find that when $\epsilon$ is constant, the 
variation vanishes.  Again the calculation parallels that done in the 
$c$-number Lagrangian case, except that the trilinear cyclic identities of 
Eq. (1c) are used in place of commutativity/anticommutativity of the fields, 
and the vanishing of terms cubic in $\chi$ is most easily established by 
using Eq.~(12a) and the cyclic identity valid for Majorana representation 
$\gamma$ matrices  (see the Appendix) 
$$\sum_{{\rm cycle}~ a \rightarrow b \rightarrow d \rightarrow a} 
(\hat\gamma^0 \gamma^{\mu})_{ab} (\hat\gamma^0 \gamma_{\mu})_{cd} 
=0~~~.\eqno(21b)$$
When $\epsilon$ is not a constant, the variation of {\bf L} is given by 
$$\eqalign{
\delta {\bf L} =&\int d^3x {\rm Tr} (\bar J^{\mu} \partial_{\mu} \epsilon) \cr
\bar J^{\mu}=&-{i \over 4g} \bar \chi \gamma^{\mu} F_{\nu \sigma}
[\gamma^{\nu},\gamma^{\sigma}]~~~,\cr
}\eqno(22a)$$
from which we construct the trace supercharge ${\bf Q}_{\alpha}$ as 
$${\bf Q}_{\alpha}=\int d^3x {\rm Tr} {i \over 8g}(p_{\chi}+i\chi^T)
F_{\nu \sigma} [\gamma^{\nu},\gamma^{\sigma}]\alpha~~~.\eqno(22b)$$
Again, it is straightforward to check, using the equations of motion and 
the cyclic identity, that ${\rm Tr} \bar J_{\mu}$ is a conserved trace 
supercurrent, which implies that the trace supercharge is conserved. 

We are now ready to check the closure of the supersymmetry algebra under 
the generalized Poisson bracket of Eq.~(5a), which for the 
Hamiltonian dynamics of the supersymmetric Yang-Mills model gives
$$
\{ {\bf Q}_{\alpha},{\bf Q}_{\beta} \}={\rm Tr}\big[ \sum_{l=1}^2
{\delta {\bf Q}_{\alpha} \over \delta A_{\ell}} 
{\delta {\bf Q}_{\beta} \over \delta p_{A_{\ell}} }
-\sum_{d=1}^4 {\delta {\bf Q}_{\alpha} \over \delta \chi^d} 
{\delta {\bf Q}_{\beta} \over\delta  p_{\chi^d}}-
\big(\alpha \leftrightarrow \beta\big) 
\big] ~~~.\eqno(23a)$$
We proceed now by the Fierz transformation method mentioned in Sec.~2 above.  
We begin by rewriting the boson terms in Eq.~(23a) as 
$$\int d^3x{\rm Tr} \sum_{\ell=1}^2 f_{\ell}^a\alpha_a\bar\beta^c g_{\ell c}
 -\big(\alpha \leftrightarrow \beta\big)~~~,\eqno(23b)$$
and the fermion terms as 
$$\int d^3x{\rm Tr} \sum_{d=1}^4  h^{da}\alpha_a\bar \beta^c k_{dc}   
-\big(\alpha \leftrightarrow \beta\big)~~~,\eqno(23c)$$
with the coefficient functions $f,g,h,k$ readily determined once the
operator variations in Eq.~(23a) have been computed.  Performing a Fierz 
transformation by using Eq.~(A.80) of West [6] then shows that verifying the 
supersymmetry algebra of Eq.~(14b), with ${\bf Q}_{\alpha,\beta}~, {\bf H}$ 
and $\vec {\bf P}$ now given by the Yang-Mills expressions of this section, 
is equivalent to verifying the three identities
$$\eqalign{
(...\gamma^0...)=&{1\over 2} {\bf H}    \cr
(...\gamma^m...)=&{1\over 2}{\bf P}^m   \cr
(....[\gamma^{\mu},\gamma^{\nu}]...)=&0~~~,\cr
}\eqno(24a)$$
with 
$$(...\Gamma...)\equiv -\int d^3x {1\over 4} {\rm Tr}
\left(\sum_{\ell=1}^2 f_{\ell}^a \Gamma_{a~}^{~d} g_{\ell d}
+\sum_{d=1}^4h^{da}\Gamma_{a~}^{~c} k_{dc} \right)~~~.\eqno(24b)$$
Equation (24b) contains both local terms, and nonlocal terms that            
couple variables at differing values of $z$.  The nonlocal terms are 
found to vanish identically when the constraint equation and symmetries 
of the integrands are taken into account, while the local terms are seen, 
by an enumeration of cases, to obey Eq.~(24a).  Examining the 
role of the supercharge as a generator of transformations, in analogy with 
Eq.~(16), in the Yang-Mills case the supercharge is found to generate 
the supersymmetry variations of Eq.~(21a), plus an infinitesimal change 
of gauge.         

We conclude by showing how the results of this section include 
the conventional case of $U(M)$ supersymmetric Yang-Mills theories, and why 
at the same time they are more general.  Consider the case in which the 
$N \times N$ matrices acted on by {\rm Tr} have dimension $N=MP$, and  
expand the matrices on a complete basis of $U(M)$ matrices $\lambda_i$, so 
that for the potential $A_{\mu}$ we have 
$$A_{\mu}=\sum_{i=0}^{M^2} {1\over 2} \lambda_i A_{\mu}^i~~~,\eqno(25a)$$
where the coefficients $A_{\mu}^i$ are now $P \times P$ matrices.  
Then the commutator term $[A_{\mu},A_{\nu}]$ in Eq.~(17b) becomes 
$$[A_{\mu},A_{\nu}]={1 \over 2} \sum_{ij} \left( [\lambda_i,\lambda_j]
\{A_{\mu}^i,A_{\nu}^j\} +\{\lambda_i,\lambda_j\}[A_{\mu}^i,A_{\nu}^j]\right)
~~~.\eqno(25b)$$
When $P=1$, so that the $A_{\mu}^i$ all commute, the second term on the 
right hand side of Eq.~(25b) vanishes, and it reduces to the conventional 
expression for the commutator term in a Yang-Mills theory.  However, our 
formalism generalizes this conventional model to allow any $P>1$, 
including the limit $P \to \infty$, in which case the 
second term in Eq.~(25b) contributes as well as the first.  Similar 
remarks apply to the other matrix commutators appearing in the derivations 
of this section.

\bigskip
\centerline{\bf 4.~~Superspace Considerations and Discussion}
\bigskip

The derivations of Secs.~2 and 3 have all been carried out in the component 
formalism, which requires doing a separate computation for each Poincar\'e 
supersymmetry multiplet.  However, there is a simple and general superspace 
argument for the results we have obtained.  Recall that superspace is 
constructed by introducing four fermionic coordinates $\theta_{\alpha}$  
corresponding to the four space-time coordinates $x_{\mu}$.  The graded 
Poincar\'e algebra is then represented by differential operators 
constructed from the superspace coordinates, and superfields are represented 
by finite polynomials in the fermionic coordinates $\theta_{\alpha}$, with 
coefficient functions that depend on $x_{\mu}$.  To generalize the 
superspace formulation to give trace dynamics models, one simply replaces 
these coefficient functions by $N\times N$ matrices (or operators), and 
one inserts a trace Tr acting on the superspace integrals used to 
form the action.  Then the standard argument that the 
action is invariant under superspace translations still holds for the trace 
action formed this way from the matrix components of the superfields.  
We immediately see from this argument why it is essential for the 
supersymmetry parameter $\epsilon$ to be a $c$-number and not also a matrix; 
this parameter appears as the magnitude of an infinitesimal superspace 
translation, and since the superspace coordinates $x_{\mu}$ and 
$\theta_{\alpha}$ are $c$-numbers, the parameter $\epsilon$ must be one also.  

The construction just given gives reducible supersymmetry representations, 
and various constraints must be applied to the superfields to pick out 
irreducible representations.  Since these constraints act linearly on the 
expansion coefficients, they can all be immediately generalized (with the 
usual replacement of complex conjugation for $c$-numbers by the adjoint) to 
the case in which the coefficient functions are matrices or operators.  

The simplicity of this argument suggests that for all nonextended rigid 
supersymmetry theories for which there exists a superspace construction,  
there should exist trace dynamics generalizations, with component  
field forms analogous to those presented above and with 
corresponding representation covariant $\gamma$ matrix identities.  
In this paper we have 
not dealt with either extended supersymmetries, or with locally 
supersymmetric theories; these will be the subject of further investigations 
into supersymmetric trace dynamics theories. 
\bigskip                                                           
\centerline{\bf Acknowledgments}
This work was supported in part by the Department of Energy under
Grant \#DE--FG02--90ER40542.  I wish to thank Ed Witten, Andrew Millard, 
and members of the Princeton graduate student supersymmetry discussion group,
for useful conversations.  
\bigskip
\centerline{\bf Appendix:  Gamma Matrix Conventions and Identities}  
\bigskip
We work with Majorana representation $\gamma$ matrices constructed 
explicitly as follows.  Let $\sigma_{1,2,3}$ and $\tau_{1,2,3}$ be two 
independent sets of Pauli spin matrices; then we take 
$$\eqalign{
\gamma^0=&-\gamma_0=-i\sigma_2 \tau_1 \cr
\hat \gamma^0=&i \gamma^0=\sigma_2 \tau_1 \cr
\gamma^1=&\gamma_1=\sigma_3 \cr 
\gamma^2=&\gamma_2=-\sigma_2\tau_2  \cr
\gamma^3=&\gamma_3=-\sigma_1 \cr
\gamma_5=&i\gamma^1\gamma^2\gamma^3\gamma^0=-\sigma_2\tau_3 \cr
\hat \gamma^0 \gamma_5=& i\tau_2 ~~~, \cr
}\eqno(A.1a)$$
so that $\hat\gamma^0,\gamma_5,\hat\gamma^0\gamma_5$ are skew symmetric 
and $\gamma^1,\gamma^2,\gamma^3$ are symmetric, and 
$$  \hat \gamma^0 \gamma^{\mu T} \hat \gamma^0=-\gamma^{\mu}  ~~~.
\eqno(A.1b)$$
For this choice of $\gamma$ matrices, the four matrices $\gamma^{\mu}$ are 
real.  

The identities of Eqs.~(12b), (15), and (21b) are easily verified by the 
following method.  Replace each four component spinor index $a,b,c,d$ by 
a pair of two component spinor indices $AA^{\prime},BB^{\prime},CC^{\prime},
DD^{\prime}$, with the unprimed indices on the matrices $\sigma$ 
and the primed indices on the matrices $\tau$.  Then each identity takes the
form
$$\Delta_{AA^{\prime}BB^{\prime}CC^{\prime}DD^{\prime}}=0~~~,\eqno(A.2a)$$
with Eq.~(A.2) symmetric under the simultaneous interchange 
$$A \leftrightarrow B,~~~A^{\prime} \leftrightarrow B^{\prime} 
~~~,\eqno(A.2b)$$
for the identities 
of Eq.~(15) and (21b) (which are symmetric under $a \leftrightarrow b$) and 
antisymmetric under this interchange for the identity of Eq.~(12b) 
(which is antisymmetric under $a \leftrightarrow b$). To verify an identity,  
it suffices to verify the vanishing of its contraction with a complete 
set of 16 projectors on the $4\times 4$ matrix with indices $a,b$, which 
in terms of $\sigma$ and $\tau$ are 
$$[\delta_{AB},(\sigma_{1,2,3})_{AB}] \otimes [\delta_{A^{\prime}B^{\prime}},
(\tau_{1,2,3})_{A^{\prime},B^{\prime}}]~~~,\eqno(A.3)$$
10 of which are symmetric and 6 of which are antisymmetric 
under the interchange of Eq.~(A.2b).  Thus 10 contractions must be done 
to verify the identities of Eqs. (15) and (21b), and 6 contractions to verify 
the identity of Eq.~(12b); these are readily done 
since the projectors involve two factors which repeat in different 
combinations, and since the contractions for individual factors 
involve only Pauli matrix arithmetic. 

The identities of Eqs. (12b), (15), and (21b) are representation 
covariant, in that they do not take the same form in representations in 
which the Dirac gamma matrices are complex rather than real.  To see this, 
we note that the matrices in a general representation $\gamma^{\mu}_G$ 
are related to the Majorana representation matrices $\gamma^{\mu}$ 
given above by
$$\gamma^{\mu}_G=U^{\dagger} \gamma^{\mu} U=U^{T*} \gamma^{\mu} U 
~~~,\eqno(A.4)$$
with $U$ a unitary matrix which in general is complex, as a result of   
which the row and column indices transform with different matrices.  
However, the identities used in the text mix row and column indices; for  
example, in Eq.~(12b) there is one term in the cyclic sum in which $a$ is 
a row index, and two terms in which $a$ is a column index.  (By way of 
contrast, the more familiar Fierz identities only interchange two row 
indices, and so do not mix row and column indices.)  Hence we cannot get 
a representation invariant form of the identity by two applications of 
Eq.~(A.4), since in the second and third terms of the cyclic sum, we 
will have a row index contracted with a $U$ and a column index contracted 
with a $U^*$, which does not correspond with Eq.~(A4).  However,   
we can easily get a representation covariant form of Eq.~(12b) by contracting 
all indices with a $U^*$, and wherever $U^*$ contracts with a column 
index using the identity 
$$U^*=UU^{*T}U^* =U\gamma^*~~~,\gamma \equiv U^TU ~~~,\eqno(A.5)$$
with $\gamma$ a matrix which appears on pp. 341-342 of the book cited in 
Ref.~[1] (because it plays a role in the transformation properties of the 
Dirac equation in quaternionic quantum mechanics).  
We can then apply Eq.~(A.4) to all the gamma matrices, giving 
for Eq.~(12b), for example, the representation covariant form 
$$\sum_{{\rm cycle}~a \rightarrow b \rightarrow d \rightarrow a} 
[(\hat \gamma^0\gamma^*)_{ab}  (\hat \gamma^0\gamma^*)_{cd}
+(\hat \gamma^0\gamma_5\gamma^*)_{ab}(\hat \gamma^0\gamma_5\gamma^*)_{cd}]
=0~~~.\eqno(A.6)$$
For a change of representation which preserves reality of the $\gamma$ 
matrices, we have $U^*=U,~~~\gamma=U^TU=U^{*T}U=1$, and 
Eq.~(A.6) is identical to Eq.~(12b), but for general changes of representation 

the identity is form covariant but not form invariant.  

\vfill\eject
\centerline{\bf References}
\bigskip
\noindent
\item{[1]}  S. L. Adler, Nucl. Phys. B 415 (1994) 195; S. L. Adler, 
``Quaternionic Quantum Mechanics and Quantum Fields,'' Sects. 13.5-7 and 
App. A (Oxford Univ. Press, New York, 1995).
\bigskip 
\noindent
\item{[2]}  S. L. Adler, G. V. Bhanot, and J. D. Weckel, J. Math. Phys. 
35 (1994), 531; S. L. Adler and Y.-S. Wu, Phys. Rev. D49 (1994) 6705.  
\bigskip
\noindent
\item{[3]}  A. C. Millard, Princeton University PhD thesis (in preparation).
\bigskip
\noindent
\item{[4]}  S. L. Adler and A. C. Millard, Nucl. Phys. B 473 (1996) 199. 
\bigskip
\noindent
\item{[5]}  S. L. Adler and   L. Horwitz, J. Math. Phys. 37 (1996) 5429.
\bigskip
\noindent
\item{[6]}  P. West, ``Introduction to Supersymmetry and Supergravity'', 
extended second ed. (World Scientific, Singapore, 1990).
\bigskip
\noindent
\item{[7]} S. L. Adler, ``The Matrix Model for M Theory as an Exemplar 
of Trace (or Generalized Quantum) Dynamics'', 
hep-th/9703053, submitted to Phys. Lett. 
\noindent
\bigskip
\noindent
\bigskip
\noindent
\bigskip
\noindent
\bigskip
\noindent
\bigskip
\noindent
\bigskip
\noindent
\bigskip
\noindent
\bigskip
\noindent
\bigskip
\noindent
\bigskip
\noindent
\bigskip
\noindent
\bigskip
\noindent
\bigskip
\noindent
\vfill
\eject
\bigskip
\bye